\documentclass[11pt]{article}

\usepackage{a4}
\usepackage[utf8]{inputenc}

\usepackage{amssymb}
\usepackage{hyperref}

\usepackage{tikz}
\usetikzlibrary{backgrounds}
\tikzstyle{every picture}=[level distance = 8mm, baseline=-0.5ex]
\tikzstyle{prop}=[shape=circle,minimum size=6mm, draw=black!80, fill=green!30]

\def\ci{{\mathcal I}}
\newcommand\In {{\mathbf I}}
\newcommand\tr {\mathop\mathbf {tr}}
\def\dirac {\slash \mkern -10mu}
\def \dirp {\slash \mkern -10mu p}

\begin{document}

\title{Higher loop corrections to a Schwinger--Dyson equation. }
\author{Marc~P.~Bellon${}^{1,2}$ and Fidel~A.~Schaposnik$^3$\thanks{Associated with CICBA}\\
\normalsize \it ${}^1$UPMC Univ Paris 06, UMR 7589, LPTHE, F-75005, Paris, France\\
\normalsize \it $^2$CNRS, UMR 7589, LPTHE, F-75005, Paris, France\\
\normalsize \it $^3$Departamento de F\'\i sica, Universidad Nacional de La Plata\\
\normalsize\it C.C. 67, 1900 La Plata, Argentina}
\date{}
\maketitle
\begin{abstract}
We consider the effects of higherloop corrections to a Schwinger--Dyson equation for propagators. This is made possible by the efficiency of the methods we developed in preceding works, still using the supersymmetric Wess--Zumino model as a laboratory. We obtain the dominant contributions of the three and four loop primitive divergences at high order in perturbation theory, without the need for their full evaluations. Our main conclusion is that the asymptotic behavior of the perturbative series of the renormalization function remains unchanged, and we conjecture that this will remain the case for all finite order corrections.
\end{abstract}



\section{Introduction}\label{int}
In preceding works~\cite{BeSc08,Be10,Be10a}, we have explored the solution of a simple Schwinger--Dyson equation for propagators, taking as a laboratory the supersymmetric Wess--Zumino model. In this way, the perturbative series of the renormalization group functions could be computed up to the high order of 200 loops and its asymptotic behavior predicted. However, these results have been obtained for a Schwinger--Dyson equation which includes only the one loop primitive divergence: the full Schwinger--Dyson equation includes an infinite number of higher loop primitive divergences. We here explore the effect of the first additional terms of the Schwinger--Dyson equation in the supersymmetric Wess--Zumino model: the three-loop and the four-loop terms (there are no primitive two-loop divergences in this model). This is by no means a simple task, since we cannot be content with the computation of the diagrams, but need to obtain a Mellin transform with eight variables in the three-loop case and eleven in the four-loop case. Furthermore, even if there is only one diagram at the level of supergraphs for each of these orders, they correspond to a sum of differing terms with possibly fermions. A scalar theory would present its own difficulties, since in this case, there will be vertex corrections and overlapping divergences which present conceptual problems.

We can however obtain valuable answers, through a chain of simple steps. The evaluation of the Mellin transform can be made with the scale invariant vacuum diagrams obtained by connecting the two external legs of the diagrams. The high degree of symmetry of these diagrams, the \(K_{3,3}\) complete bipartite diagram at three loops and the cube at four loops, simplify many computations. We concentrate on terms which have the potential to change the asymptotic behavior of the perturbative series, which is not the case for single terms in the Taylor expansion of the Mellin transform. The dominant contributions therefore stem from the leading parts of the asymptotic expansion of the Taylor series, which can be deduced from the singularities near the origin of the Mellin transform. These singularities in turn stem from the divergences which appear when some subdiagrams become scale invariant. The poles are easy to determine and the values of the residues which dominate our computation reduce to known bosonic integrals.

We shall first recall some basic points of our precedent works, which motivate the present study. We next present the link between the potential divergences of subdiagrams and the poles of the Mellin transform, allowing a complete study of the associated residues. For simplicity, the presentation is made for scalar diagrams, but the possible numerators, coming from fermion contributions in our case, add only inessential computational complexities. We then consider the numerators for the combination of diagrams imposed by supersymmetry. Their full evaluation in a parametric representation of the diagrams is beyond the possibilities of the classical methods, if we want to keep the high degree of symmetry which one expects: we therefore limit ourselves to the cases which are necessary for the evaluation of the dominant residues. We then combine all these results to obtain the leading asymptotic contributions of these two diagrams. 

\section{Summary of previous work.}

Our solution strategy for the Schwinger--Dyson equation was developed in former works~\cite{BeSc08,Be10}, based on the proposal of~\cite{KrYe2006}. A first ingredient is a renormalization group argument which allows us to express a full propagator as a series in the logarithm of the norm of the momentum from the renormalization group functions for a minimal computational cost. With the power of methods using Hopf algebras of graphs~\cite{Connes:1999zw,Connes:1999yr,CoKr00} we showed that these renormalization group arguments apply to the solutions of truncated Schwinger--Dyson equations. It has further been remarked in~\cite{Be10a} that not only the different derivatives of the propagator can be computed, but also some infinite sums of derivatives.

In this simple case with only renormalized propagators, the remaining problem is the evaluation of primitively divergent diagrams with full perturbative propagators. Since perturbative propagators are polynomials in the logarithm of the squared momentum, all such evaluations can be derived from a  single analytic function, with parameters appearing as exponents of the different propagators of the diagram. A Taylor expansion then gives the effect of any powers of the logarithms of the propagators. The use of such parameters is the basis of the analytic regularization which enjoyed a brief popularity before being superseded by dimensional regularization~\cite{BoGi64,Sp69,Sp71}.  Such Mellin transforms now have a rich history as a tool for the evaluation of Feynman diagrams: in particular, Mellin transforms are instrumental in~\cite{BiWe2003} to compute multiloop diagrams from single loop parts.

What will be of tremendous use in what follows is that the Mellin transform, which is defined a priori as an analytic function on a domain of convergence depending only on the real parts of the parameters, can be extended on the whole complex domain as a meromorphic function. Its poles are linked to divergences, when subdiagrams become scale invariant. In a parametric integral representation of these Mellin transforms, factors of the inverse \(\Gamma\) function appear, which give a number of zeros. For even integer dimensions and integer values of a number of the parameters, these zeros cancel but a finite number of poles, so that one obtains a rational function. Such simplifications will allow us to obtain the dominant contribution from a priori very complicated diagrams.

\section{Scale invariant diagrams.}
\label{sclint}

We begin by the study of bosonic diagrams, avoiding the complexities stemming from numerators in the supersymmetric case. We start from a diagram with two external legs \(G\), to be evaluated with an external momentum \(p\), with a Mellin transform variable \(\beta_i\) associated to each edge\footnote{Physicist are used to calling them lines, but we shall stick to the terminology of graph theory. Line will be used to denote a path formed by Fermionic propagators. In the supersymmetric cases, we have to sum on the different configuration of Fermionic lines.}.

It is convenient to use a parametric representation of the diagram \(G\) of loop number \(L\), with Schwinger parameters \(a_i\), integrated from 0 to \(\infty\), for each edge \(i\). The first and second graph polynomials \(U\) and \(V\), respectively of degrees \(L\) and \(L+1\), give a simple expression of the integral on loop momenta in dimension \(D\):
\begin{equation}
  	\In[\beta_i] = \frac 1{\prod \Gamma(\beta_i) } 
		\int \prod \Bigl(da_i^{\vphantom {\beta_i}} a_i^{\beta_i-1}\Bigr) U^{-D/2}
		 \exp\left({-\frac V U p^2}\right).
\end{equation}
Usually one simply integrates on the overall scale of the \(a_i\), but we can introduce an additional variable \(a_0\) to express the product of powers of \(U\) and \(V\)  in terms of \(\tilde U = V + a_0 U\), which is the polynomial associated to the vacuum diagram \(\tilde G\) obtained by completing the original diagram with an edge joining the two vertices linked to the exterior. With the exponent \(\beta_0\) such that the sum of the \(\beta_i\) is \((L+1)D/2\) for a diagram with \(L\) loops, \(\In [\beta_i]\) takes the following very symmetric form, depending only on \(\tilde G\), apart from the dependence on the exterior momentum:
\begin{equation} \label{InS}
	\In_{\tilde G} [\beta_i] = (p^2)^{\beta_0 - D/2} \frac {\Gamma(D/2)} {\prod \Gamma(\beta_i)} \int \prod \Bigl( da_i^{\vphantom {\beta_i}} a_i^{\beta_i-1}\Bigr) \tilde U^{-D/2} \delta (\textstyle{ \sum' }a_i -1),
\end{equation}
where now the products include factors with index \(i=0\) associated to the edge closing the graph. The values of the \(\beta_i\) give an overall scale invariance to the integrand in this formula: the \(\delta\) factor breaks this scale invariance, avoiding an infinite factor. The sum is primed in this \(\delta\) factor to indicate that any subset of the \(a_i\) can be included in this sum, without changing the value of the integral. This freedom makes the analysis of the singularities of this integral easier. This passage from \(G\) to its completed graph \(\tilde G\) is not a new remark: it is the first step in the elucidation of the invariance of the irreducible two-loop propagator graph under the \(\sigma_6\) group~\cite{GoIs84,Br86}. Its generalization to arbitrary graphs has been invoked in the classification of \(\phi^4\) transcendentals~\cite{Sch10}.

Consider indeed any 1PI subdiagram \(K\) of \(G\) and the new diagram \(G/K\) defined through the contraction of the subdiagram \(K\), i.e., deleting the edges of \(G\) belonging to \(K\) and identifying their ends. If one gives weights 1 to the edges of \(K\) and weights 0 to those of \(G/K\), the lowest weight terms of \(U_G\) factorize into \(U_{G/K} U_K\)~\cite{Na71}. If we introduce a new variable \(a\) appearing as a factor in all the variables associated to \(K\), the integral in equation~(\ref{InS}) becomes, if \(K\) has \(\ell\) loops,
\begin{equation}
  \int \prod \Bigl( da^{\vphantom {\beta_i}}_i \, a_i^{\beta_i-1}\Bigr) \delta (\textstyle{ \sum' }a_{i'} -1) \delta (\textstyle{ \sum'}a_{i''} -1)
  	\int da \, a^{\beta -1} (a ^ {\ell} U_{G/K} U_K + \cdots)^{-D/2},
\end{equation}
with \(\beta = \sum_{i'\in K} \beta_{i'}\). The indexes \(i'\) are associated to the edges in \(K\) and the \(i''\) to the edges in \(G/K\). The second \(\delta\) is there to compensate the introduction of the additional integration variable \(a\).  The \(a\) integral is convergent at 0 when \(\beta - \ell D/2\) is positive and can be extended as a meromorphic function with simple poles when \(\beta-\ell D/2\) is 0 or a negative integer, with residues proportional to the integrand and its derivatives at \(a=0\). For \(\beta - \ell D/2 = 0\), the integrations over the \(a_i\) associated to \(K\) and \(G/K\) factorize, so that the residue of \(\In[\beta_i]\) is given by
\begin{equation}
 \mathrm{Res}_{\beta - \ell D/2 = 0} \,\In[\beta_i] = 
 1/\Gamma(D/2)\In_K[\beta_{i'}] \In_{G/K}[\beta_{i''}] \; ,
\end{equation}
The condition on \(\beta\) is exactly what is necessary for both \(G/K\) and \(K\) to be scale invariant, so that the integrals are well defined. For the further poles \(\beta - \ell D/2 = -k\), the denominator is still factorized, of the form \(U_K^{-D/2-l} U_{G/K}^{-D/2-l}\), with \(l\) an integer less or equal to \(k\). The residue therefore is a sum of factorized integrals with polynomial numerators. Such numerators, as well as the ones coming from the consideration of non-scalar diagrams will not change the type of poles we can expect in these residues: they can merely shift them by integer values.

We conjecture that these poles are the only singularities of the meromorphic extension of \(\In_G[\beta_i]\) with double and higher poles associated to chains of inclusions of diagrams \(G_1 \subset G_2 \subset \cdots \subset G_l = G\). For a \(L\) loop diagram, the poles of order \(L-1\) come from chains of inclusions such that each quotient \(G_{i+1}/G_i\) is a one loop diagram: the corresponding integrals will be of the form
\begin{equation} \label{1loop}
 	\int \prod  \Bigl( da^{\vphantom {\beta_i}}_i \, a_i^{\beta_i-1}\Bigr)
		P(a_i) (\sum a_i)^{-D/2-l} \delta(a_{i_0}-1),
\end{equation}
with \(P\) a polynomial whose degree is completely fixed by the requirement of scale invariance. For each monomial of \(P\), this generalized \(B\) function evaluates to a product of Euler \(\Gamma\) functions of arguments \(\beta_i\) shifted by integers, divided by \(\Gamma(-D/2-l)\). Divided by the product of \(\Gamma(\beta_i)\) in the definition of \(\In\), this simplifies to polynomials in the \(\beta_i\), with rational coefficients when \(D/2\) is an integer. If there are no numerators, as is the case for the poles at the limit of the convergence domain, the residue is constant.

The core Hopf algebra introduced in~\cite{Kreimer:2009jt,Kreimer:2009iy} could help control the combinatorics associated to the different decompositions in chain of subdiagrams. This algebra is similar to the Connes--Kreimer Hopf algebras of diagrams~\cite{Connes:1999yr}, but without the restriction to diverging subdiagrams in the left part of the coproduct. The quotient diagram in the right part of the coproduct will therefore have vertices with any valence. Indeed, with a proper choice of the variables \(\beta_i\), any subdiagram can become divergent. The order \(n\) poles of \(\In[\beta_i]\) should be read from \(\bar\Delta^n G\).\footnote{\(\bar\Delta\) is used to denote the reduced coproduct, which is obtained by eliminating the trivial \(1\otimes G\) and \(G \otimes 1\) terms in the coproduct.} However the pole structure is not directly linked with the different terms in such coproducts. Whenever a factor of the coproduct correspond to a product of independently homogeneous terms, the residue of the pole is zero. This can happen when a factor is either not connected or one vertex reducible, {\em i.e.}, which can be disconnected by removing a single vertex. Finally, different terms of \(\bar\Delta^n G\) could be formed of the same subdiagrams in differing orders: each of these terms  gives the same poles, which should probably appear only once in the final result. Devising an efficient may of organizing these poles and computing the residues is therefore beyond the reach of this work.

\section{Supersymmetry and numerators.}
As advanced in the introduction, we intend to consider a supersymmetric Wess-Zumino model defined in terms of a single chiral  superfield
composed of a complex scalar, a chiral fermion and a complex auxiliary field. At the order of our computations, the only consequence of the chirality of the fermion field is to divide by 2 the trace over spinor indices involved in fermion loops. 
Supersymmetry introduces both simplifications and complexities. The presence of fermions introduces numerators in the Feynman integrals. The special combinations suppress a number of divergences, making the whole subdivergence structure simpler. On the other hand, these numerators which result from the combinations of a large number of terms have no simple general expression. In the cases of the three- and four-loops corrections however, the simplest divergences are given by purely bosonic diagrams.

The only primitive three-loop correction in the Wess--Zumino model is given by the following non-planar diagram:
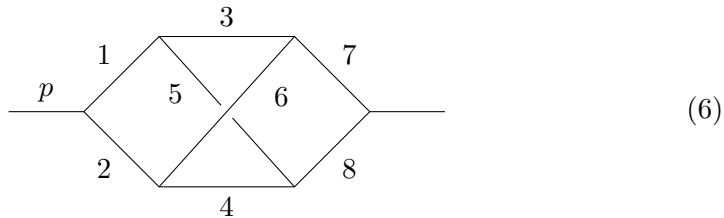
\begin{equation} \label{diag3L}
\begin{tikzpicture}
	\draw (0,0) -- node[above](){\(p\)}++(1,0) coordinate(entree){} --node[above left](){1} ++(1,1) coordinate(ul){} -- node[above](){3} ++(1.8,0) coordinate(ur){}
		--node[above right](){7}  ++(1,-1) coordinate(sortie) -- ++(1,0) ;
	\draw (entree) -- node[below left](){2} ++(1,-1) coordinate(dl){} --node[below](){4} ++(1.8,0) coordinate(dr){} --node[below right](){8} (sortie);
	\begin{pgfonlayer}{background}
	\draw (ul) --node[below left, near start](){5} (dr); \draw[draw=white, double=black, double distance= 0.4pt, line width=3pt] (ur)--node[below right, near start](){6}(dl);
	\end{pgfonlayer}
\end{tikzpicture}
\end{equation}

Indeed, for zero external momentum in the auxiliary field propagator, the combination of numerators becomes simply the product of the squared momenta on edges 1 and 7, so that the residue of the diagram is given by the tetrahedron diagram with exponents \((\beta_1+\beta_2 -1 , \beta_3, \beta_4, \beta_5, \beta_6, \beta_7+\beta_8-1)\). The same result can be obtained for the fermionic and bosonic propagators, but the necessary algebra is better done with computer assistance: we used Form from J.A.M.~Vermaseren~\cite{Ver2000}.

The first infrared divergences for any internal edge are likewise given by purely bosonic diagrams. The first infrared divergence indeed comes from an edge with a boson propagator, since the numerator in either fermionic or auxiliary propagators would lower the infrared divergence degree. This can entirely fix the particle assignments on all edges, when we consider the edge 3 in the propagator for the auxiliary field, or strongly constrain the possibilities. In any case, the two edges meeting a zero momentum edge of bosonic character will share the same momentum \(q\): either one of them is an auxiliary field which contribute a numerator \(q^2\) or the two are fermionic lines with numerator \(\dirac q \; \dirac q = q^2\). In any cases, the two lines combine to a single bosonic line with exponent \(\beta + \beta'-1\). 

For the divergence associated to \(\beta_3 = 2\), we simply obtain in the auxiliary field case the two-loop diagram with exponents \((\beta_1+\beta_5-1, \beta_2, \beta_4, \beta_8, \beta_6+\beta_7-1)\). This can be completed to a tetrahedral scale invariant diagram with an edge of degree \(\beta_0\) such that the sum of the \(\beta_i\) is \(10 = 4 D/2 + 2\). 
For the case of the fermion propagator, only three of the eight possible configurations survive when we impose that edge 3 is a boson propagator and they yield respectively factors of \(\dirp_2\), \( -\dirp_4\)  and\(\dirp_8\) apart from the \(p_5^2\, p_6^2\) factor common to all configurations (all momenta are supposed to go from left to right). These factors add up to \(\dirp\), so that supersymmetry is verified and the divergence is given by the same bosonic diagram as for the auxiliary field. In the bosonic cases, there are six contributions: the diagrams with only auxiliary and bosonic fields yield factors \(p_2^2\), \(p_4^2\) and \(p_8^2\) while the fermionic loops yield factors of \(2 \,p_2\cdot p_8\), \( -2 \,p_2\cdot p_4\) and \(-2 \,  p_4\cdot p_8\) which add up to \(p^2\), again trivially verifying supersymmetry. Combined with the symmetries of the diagram, this solves the cases of all the edges which do not meet the exterior ones. 

In the case of the divergence on the edge 1 for the propagator of the auxiliary field, the three possible field configurations contribute and they give factors \( p_5^2 \, p_6^2\), \(p_5^2 \, p_4^2\) and \( 2 p_5^2 \, p_4\cdot p_6\) for the fermionic loop. They add  up to \(p_5^2 \,p^2\). In the cases of fermionic and bosonic terms, it is easy to see that one obtains the same numerator, multiplied either by \(\dirac p\) or \(p^2\). The strong symmetry between the different positions of the divergence suggests that, as in the purely bosonic case, a highly symmetric formulation is possible, based on the complete bipartite graph \(K_{3,3}\) obtained by closing the graph. In every cases, the divergence is given by the residue of the wheel with three spokes, which is known to be proportional to \(\zeta(3)\).

We now turn to the more challenging case of the four loop graph
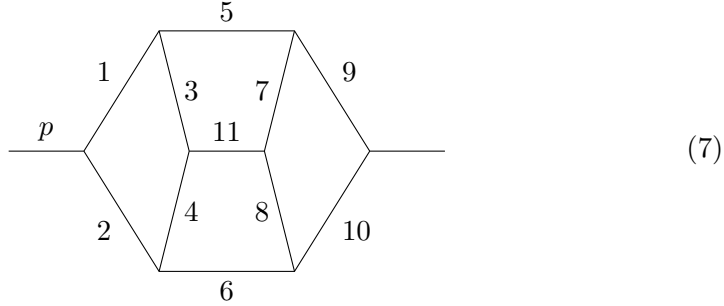
\begin{equation} \label{diag4L}
\begin{tikzpicture}
	\draw (0,0) -- node[above](){\(p\)}++(1,0) coordinate(entree){} --node[above left](){1} ++(1,1.6) coordinate(ul){} -- node[above](){5} ++(1.8,0) coordinate(ur){}
		--node[above right](){9}  ++(1,-1.6) coordinate(sortie) -- ++(1,0) ;
	\draw (entree) -- node[below left](){2} ++(1,-1.6) coordinate(dl){} --node[below](){6} ++(1.8,0) coordinate(dr){} --node[below right](){10} (sortie);
	\draw (ul) -- node[right]{3}+(.4,-1.6) coordinate(ml){}-- node[right]{4}(dl);
	\draw (ur) -- node[left]{7}+(-.4,-1.6) coordinate(mr){}-- node[left]{8}(dr);
	\draw (ml)--node[above]{11} (mr);
\end{tikzpicture}
\end{equation}
The simplest situation is when we consider a divergence on the edge 5 for the auxiliary field: then the propagators 3 and 7 must be auxiliary fields and the numerator is completely fixed with the auxiliary field in 6. Apart from the factors coming from the neighbors of 5, we have the single factor \(p_6^2\). The cases of fermionic and bosonic propagators are more complex but can be easily solved. In the fermionic case, the fermion line can go either through 11 or 6. In the former case, there is only one possibility, which gives the numerator \(p_6^2 \,\dirac p_{11}\). In the latter case, one obtains \( (\dirac p_2 + \dirac p_4) \dirac p_6 ( \dirac p_8 + \dirac p_{10} ) \), which gives \( \dirac p_6 \,\dirac p_6\,\dirac p_6 = p_6^2\, \dirac p_6\). The two add to form the result predicted by supersymmetry, \(p_6^2 \,\dirac p\). Likewise, one can show that for the bosonic propagator, one obtains a term \(p_{11}^2\, p_6^2 \) when 11 is an auxiliary field, \(2\, p_{11}\cdot p_6\,p_6^2\) when 11 is fermionic and \(p_6^2 \, p_6^2\) in the last case where 11 is bosonic, for a total which is \(p_6^2\, p^2\).

With the infrared divergence on the edge 11, one obtains \(p_5^2 + 2\, p_5\cdot p_6 + p_6^2 = p^2\) with the auxiliary field, \( (\dirp_1 + \dirp_2) ( \dirp_5 +\dirp_6) (\dirp_9 + \dirp _{10}) = \dirp \,\dirp \,\dirp\) in the fermionic case and \( ( p_1^2 + p_2^2 + 2 p_1\cdot p_2)( p_1^9 + p_{10}^2 + 2 p_9\cdot p_{10}) = \bigl(p^2\bigr)^2 \). Again, supersymmetry is simply verified. One remarks that the overall divergence degree is lowered. For a divergence on edge 3, one obtains a numerator proportional to \(p_{10}^2\) and again supersymmetry relations.

Let us detail the remaining case of a divergence on edge 1, which subsumes all necessary procedures. In the case of the auxiliary field, we must separately consider the cases where the propagator 6 correspond to the three possible field types. With a boson propagator, all other propagators are fixed and we get \( p_4^2 \,p_8^2 \) beyond the common \(p_3^2\). With an auxiliary field on edge 6, three configurations are possible, which give terms \(p_6^2 \,p_{11}^2\), \(p_6^2 \,p_7^2\) and \(2 p_6^2 \, p_7\cdot p_{11} \), adding up to \(p_6^2 \, p_8^2\). Finally, if the propagator 6 is fermionic, the Fermion line can close either through edge 5 or 11, giving \(\tr( \dirp_{11} \dirp_8 \dirp_6 \dirp_4) \) and \(\tr( \dirp_7 \dirp_8 \dirp_6 \dirp_4) \). Now, \(p_7+p_{11} = p_8\), so that these two terms give \(2 p_4\cdot p_6 p_8^2\). This three different terms have \(p_8^2\) as a common factor, and the remaining factors add to \(p_2^2 = p^2\). 

In the case of the fermionic propagator, the fermionic line goes first through edge 2 and we have four possibilities according to the next step being edge 4 or 6 and the final one being edge 9 or 10. In each of these cases, a factor \(p_8^2\) appears, either directly when edge 8 is an auxiliary field propagator, or through the combination \(\dirp_8 (\dirp_7 + \dirp_{11})\) or through the combination of the fermionic loop \((3,5,7,11\) and the two possible auxiliary field positions on this loop. One therefore obtains \(p_8^2 \dirp_2 (\dirp_4 + \dirp_6) ( \dirp_9 + \dirp_{10}) = p_8^2 p^2 \dirp\). In the case of the bosonic propagator, a factor \(p_2^2 = p^2\) comes from the edge 2 which must be an auxiliary field and the three possibilities, auxiliary field on 9, auxiliary field on 10 and fermionic line going through both of them, have a \(p_8^2\) factor. The additional factors are \(p_9^2\), \(p_{10}^2\) and \(2 p_9\cdot p_{10}\) which add up to \(p^2\). Therefore, also in this case is supersymmetry simply verified and the divergence is given by a simple bosonic diagram with exponents \((\beta_8-1,\beta_3+\beta_5-1,\beta_2+\beta_0-1,\beta_i)\).

Finally, we must consider the overall divergence of the diagram, which can be inferred in the limit of a vanishing exterior momentum. In the case of the auxiliary field propagator, it can be shown that the different numerators combine to give the product \(p_{11}^2 p_1^2 p_9^2\), so that the residue is given by a scalar diagram with exponents \((\beta_1 + \beta_2-1,\beta_9+ \beta_{10}-1, \beta_{11}-1)\), with all other exponents unchanged. The supersymmetry relations become harder to verify in this case, because they can no longer be verified at the level of the integrands through automated expansion. However, since the differences are proportional to the antisymmetric Levi-Civita tensor, the momentum integration is trivially zero.

All these explicit computations tend to sustain the conjecture that the supersymmetric Mellin transform can be obtained from a fully symmetric parametric integration on the completed diagram, which is a cube. In any case, the singularities are always given by the wheel with four spokes graph, a diagram whose evaluation is known to be proportional to \(\zeta(5)\) (see for instance~\cite{BrYe09}).

It should be noted that these higher loop primitive parts have contribution of higher weight than the ones obtained through the computation of our preceding works. Indeed, in~\cite{BeSc08}, the first term proportional to \(\zeta(3)\) appears at order four and the one proportional to \(\zeta(5)\) at order six, at respectively one and two orders beyond the first contributions of the three and four loop primitives. This is consistent with the phenomenon of weight drop for diagrams with subdivergences described in~\cite{BrScYe12}.

\section{Applications.}

The cases we considered in the preceding section correspond to a small part of the possible singularities, since they only are the first ones associated to subgraphs with one loop quotients as well as the value at the origin. The study of section~\ref{sclint} shows that there are further poles associated to the same subdiagrams as well as singularities associated to the subdiagrams with any number of loops and any combination of these elementary singularities. However, these additional terms will give subdominant contributions to the perturbative series of the \(\beta\) function and the limited knowledge we have will be sufficient for our purpose.

As in the fourth section of~\cite{Be10a}, where such general poles of the Mellin transform are studied, we consider the pairing as given by replacing the variables \(x_i\) such that \(\beta_i =1 -x_i\) by a derivative with respect to the logarithm of the momentum \(L_i\). If we call \(\ci(k,p)\) the object associated to the Mellin transform
\begin{equation}\label{pole}
	\frac {N(x_i)} { k + \sum_{i=1}^p  x_i},
\end{equation}
it will be given by the following expression
\begin{equation}
	\ci (k,p)= \left.\frac1{k+\sum_{j=1}^p \partial_{L_j}} N(\partial_{L_1},\ldots,\partial_{L_p}) 
		\prod_{j=1}^p G(L_j)\right|_{L_1=\cdots=L_p=0}
\end{equation}
The derivatives with respect to the \(L_j\)'s can be converted through the use of the renormalization group to the expression \(\gamma + \beta \, a \partial_a\). In the absence of vertex renormalization, \(\beta\) is simply \(3\gamma\) and the differential operator in \(a\) which is now in the denominator can be brought to the left hand side to give 
\begin{equation} \label {resume}
\bigl( k +  \gamma(p + 3 a \partial_a)\bigr) \ci (k,p) =  N'(\gamma,\gamma_2, \ldots) 
\end{equation}
with \(N'\) some function of the derivatives of \(G\) with respect to \(L\) which can be deduced from the numerator \(N\). Equation~(\ref{resume}) gives a recursion relation for calculating the power series expansion of \(\ci(k,p)\).  Whenever the absolute value of \(k\) is not 1, the terms depending on lower order terms of  \(\ci(k,p)\) are of the same order than the terms directly proportional to the \(\gamma_i\): the growth of  \(\ci(k,p)\) is therefore like the one of \(\gamma\). In the determination of \(\gamma\),  \(\ci(k,p)\) is multiplied at least by \(a^l\), with \(l\) the loop order of the correction to the Schwinger--Dyson equation, so that such contributions are smaller by a factor \(1/n^l\) at order \(n\) with respect to the principal term in \(\gamma\). 

We must therefore find all the poles with \(k = \pm 1\). The effect of the numerator in the Feynman integral on the position of the first pole depends only from its degree. With supersymmetry, this first pole can have a vanishing residue, shifting by one unit the family of poles.  

The scale invariance of quotients \(G/K\) with more than 1 edge holds for sums of the associated \(x_i\) of \(-2\) or less. For a quotient with two loops and three edges, there are only three denominators to compensate the two integrations and the numerator, which sum to 5, so that the sum of the \(x_i\) must be at least \(-2\). The quotient with three loops, five edges has at least one unit from the numerator, giving again \(3 \times 2 + 1 - 5 =2 \) to be compensated by the \(x_i\). In the case of the cube, there is also a four loop quotient, with at least two powers of \(p^2\) in the numerator, and eight edges: again the total of the \(x_i\) must be \(-2\). The only quotients giving terms with \(k=1\) are therefore the simple edges, giving terms proportional to \(\ci(1,1)\). Since \(a \ci(1,1)\) is the dominant term in the determination of \(\gamma\)~\cite{Be10a}, this will give relative corrections of order \(1/n^{l-1}\), i.e., of order \(1/n^2\) for the three-loop diagram and \(1/n^3\) for the four loop cube. Corrections of the same order in the asymptotic expansion come from the modified order 3 and order 4 values of \(\gamma\) which appear in the recursion for \(\ci(1,1)\), which is the main contributor to \(\gamma\), according to~\cite{Be10a}.

The situation for the terms with \(k=-1\) is less simple. However, there is certainly one term stemming from the divergence of the full diagram. Indeed, there is a pole when the sum of all the \(x_i\) is zero, which is compensated when taking the derivative with respect to the external momentum, but there are also poles for all the negative integer values of this sum, even if the exact numerator would be difficult to determine. Now, the number of propagators \(p\) is equal to \(3 l -1\) for the term of order \(l\) in the Schwinger--Dyson equation. If we now go for the recurrence equation for the contribution to \(\gamma\), \(a^l \ci(-1,p)\), it becomes independent on \(l\), except in the right hand side:
\begin{equation} \label{finale}
\bigl( -1 + \gamma(3a \partial_a - 1) \bigr) [a^l\ci(-1,3l-1)]= a^l N'(\gamma,\gamma_2,\ldots)
\end{equation}
The dominant non-alternating term in the perturbative series for \(\gamma\), therefore receives contributions not only from the order one term in the Schwinger--Dyson equation, but also from its corrections at all orders. In any case, the evaluation of the coefficient for this asymptotic term should be obtained by a fit to the finite number of terms of the perturbative series explicitly computed: the important result here is that such a fit should still be based on the same basic series, one with alternating and the other with non-alternating signs, and their corrections by powers of \(1/n\).

\section{Conclusion}

Our preceding works on the solution to high order of Schwinger--Dyson equations were limited to the simplest ones with only one-loop primitive divergences. The inclusion of higher-loop primitive divergences seemed dauntless. Indeed, the evaluation of the associated Feynman diagrams has to take into account logarithmic corrections of any order in each and every one of the numerous propagators in the diagram. Moreover the evaluation of individual contributions is but the beginning, since there are numerous possible contributions at a given degree in \(a\), corresponding to different ways of obtaining this total degree from the sum of the degrees of the individual propagators. Despite all these odds, we succeeded to obtain the dominant effect of such a diagram on the asymptotic behavior of the perturbative series for the \(\beta-\)function of the renormalization group. It seems even possible to derive corrections to this dominant effect.

This radical change of perspective stems from the combination of the successive progresses made in our previous works~\cite{BeSc08,Be10,Be10a} with the derivation of the general structure of the singularities of the Mellin transform of any diagram made in section~\ref{sclint}. Here are the successive steps towards this achievement: 
\begin{itemize}
\item compute the primitive diagrams with propagators raised to any powers, giving a Mellin transform from which the effects of the logarithmic corrections to the propagators can be obtained through a Taylor expansion~\cite{KrYe2006}.  
\item recognize that the asymptotic properties of the Taylor series stem from the singularities of the function, so that simple asymptotic formulas for these coefficients can be given~\cite{Be10}.
\item for poles of some simple form of the Mellin transform, use the renormalization group to compute the exact contributions stemming from the whole Taylor expansion of this pole~\cite{Be10a}. For poles with denominators involving many variables, this means that we can obtain with a simple operation of low computational cost the sum of great number of terms.
\item show that for any diagram, the Mellin transform has only poles of the type covered in the preceding item, with computable polynomial residues (this work).
\end{itemize}

In this article, this whole scheme was illustrated on two specific primitive divergences of the Wess--Zumino supersymmetric model. We did not completely solve the challenge of the evaluation of the numerators, which involves the sum of different ones coming from supersymmetry, but we could deduce those necessary for the evaluation of the dominant singularities. These partial results suggest a high level of symmetry, which should allow a complete description of the integral form of the Mellin transform and the computation of all residues: we hope to be able to present such results in future works.

Finally, through the combination of all these insights, we have been able to describe the dominant effects of the first corrections to the Schwinger--Dyson equations on the asymptotic behavior of the perturbative series. The remarkable point is that, even if we cannot compute exactly the full Mellin transforms of the primitively divergent graphs, much can be said on the order of magnitude of the ensuing corrections without any explicit computation.

A clear distinction appears between the contributions with constant signs and those with alternating signs. The latter ones are dominant in absolute values, have unambiguous sums through the Borel method and receive contributions which are suppressed by powers of the order from the higher loop primitive graphs. It therefore seems possible to have a systematic expansion of the higher order contributions, since only a finite number of primitive graphs give contributions to each order of the expansion. This could be trumped by the number of primitive divergences, which is a finite fraction of the total number of graphs, but the result should in any case hold in the large \(N\) limit where only planar graphs contribute. If we had only these terms, we could think that a constructive determination of the renormalization function is at hands. However the parts with constant sign are there and cumulate difficulties: they are associated to singularities of the Borel transform for positive arguments, implying ambiguities in the summation process, and every order of the asymptotic expansion receives contributions from all possible terms in the Schwinger--Dyson equation.

\noindent {\bf Acknowledgements}: F.A.S is associated to CICBA and partially supported by CONICET  ANPCyT  and CICBA grants.  

\bibliographystyle{unsrturl} 
\bibliography{renorm}

\end{document}